\documentclass[useAMS,usenatbib]{mn2e}
\pdfoutput=1

\usepackage{changepage}
\usepackage{booktabs}
\usepackage{aas_macros}
\usepackage{epsfig}
\usepackage{amsmath}
\usepackage{amssymb}
\usepackage{comment}
\usepackage{caption}
\usepackage[justification=centering]{caption}

\usepackage{leftidx}
\usepackage{natbib}
\usepackage[rgb]{xcolor}
\definecolor{MyGreen}{rgb}{0.0,0.6,0.3}
\definecolor{MyPurple}{rgb}{0.6,0,0.3}
\usepackage{fancyref}
\usepackage{hyperref}
\hypersetup{colorlinks=true,citecolor=MyGreen,linkcolor=MyPurple,urlcolor=blue}

\def\beq{\begin{equation}}
\def\eeq{\end{equation}}
\def\ba{\begin{eqnarray}}
\def\ea{\end{eqnarray}}
\def\bal{\begin{align}}
\def\eal{\end{align}}
\def\bxi{{\mbox{\boldmath $\xi$}}}

\begin{document}
\title[Slowly Spinning Stars] {Super Slowly Spinning Stars in Close Binaries}
\author[Fuller \& Felce]{
Jim Fuller$^{1}$\thanks{Email: jfuller@caltech.edu} and 
Catherine Felce$^{1}$
\\$^1$TAPIR, Mailcode 350-17, California Institute of Technology, Pasadena, CA 91125, USA
}
\label{firstpage}
\maketitle
\begin{abstract}

Stars in short-period binaries typically have spins that are aligned and synchronized with the orbit of their companion. In triple systems, however, the combination of spin and orbital precession can cause the star's rotation to evolve to a highly misaligned and sub-synchronous equilibrium known as a Cassini state. We identify a population of recently discovered stars that exhibit these characteristics and which are already known to have tertiary companions. These third bodies have a suitable orbital period to allow the inner binary to evolve into the sub-synchronous Cassini state, which we confirm with orbital evolution models. We also compute the expected stellar obliquity and spin period, showing that the observed rotation rates are often slower than expected from equilibrium tidal models. However, we show that tidal dissipation via inertial waves can alter the expected spin-orbit misalignment angle and rotation rate, potentially creating the very slow rotation rates in some systems. Finally, we show how additional discoveries of such systems can be used to constrain the tidal physics and orbital evolution histories of stellar systems.

\end{abstract}

\section{Introduction}

It is well known that the lowest energy state of a binary star system has stellar spins that are synchronized and aligned with the orbit \citep{Hut_1980}. Dissipative processes will eventually bring the system into such a state, as observed in many close binary systems (e.g., \citealt{Lurie_2017}). And yet, several recently discovered systems are clearly not in this state (\autoref{tab:system_params}). In fact, they have extremely slow rotation rates compared to both single stars and close binaries. 

A possible explanation is that those systems are not binaries, but rather triple systems. In triple systems, the combination of orbital precession (induced by the third body) and spin-orbit precession (induced by the central body's centrifugal bulge) changes the equilibrium spin-orbit configurations. Importantly, there arises a second stable state known as Cassini State 2 (CS2), which typically exhibits a large degree of spin-orbit misalignment and a sub-synchronous rotation rate \citep{Colombo_1966,Peale_1969}.

Recent work has investigated Cassini equilibria in the context of multi-body exoplanet systems \citep{Winn_2005,Fabrycky_2007,Storch_2017,Anderson_2018,Millholland_2018,Stephan_2018,Millholland_2019,Millholland_2020,Su_2020,Su_2021,Huang_2023}. These papers have worked out much of the relevant physics and made predictions for rotation rates and spin-orbit misalignment angles. Several of them also include complex dynamics accounting for orbital eccentricity, Kozai-Lidov cycles, and high-eccentricity migration. \cite{Anderson_2018} demonstrated how spin-orbit misalignment between a star's spin axis and an inner companion can be resonantly excited via the action of a third body. \cite{Su_2021} included the effects of tidal dissipation, showing how planets could evolve towards either stable Cassini state. Unfortunately, the spin orientation and spin rates of low-mass exoplanets are nearly impossible to measure, creating a barrier for testing such theories.

In stellar systems, however, spin rates can be measured either asteroseismically \citep{Li_2020a}, through spots \citep{Lurie_2017}, or rotational broadening of spectroscopic lines. Spin-orbit misalignment can also be measured via the Rossiter-Mclaughlin effect in eclipsing systems (e.g., \citealt{Albrecht_2009,Albrecht_2014}). Hence, stellar triples are a natural testing ground for spin-orbit dynamics, Cassini equilibria, and tidal dissipation mechnanisms. The possibility of slow and misaligned stellar binaries caught in Cassini equilibria has been recognized \citep{Naoz_2014,rose:19} but has not been widely appreciated.

In this paper, we extend previous planetary work to the case of stellar triples, revealing the parameter space over which these systems can be caught in CS2, and computing the corresponding spins and obliquities. We apply these calculations to well characterized stellar triple systems, finding that some of them have rotation rates even slower than expected. We investigate tidal dissipation via inertial waves as a possible explanation for the very slow rotation, and we comment on some of the other important physical mechanisms applicable to stellar systems that have not been investigated in prior planet-focused work.

\section{Spin and Orbital Dynamics}
\label{sec:spin}

\begin{figure}
\centering
\includegraphics[width=0.49\textwidth]{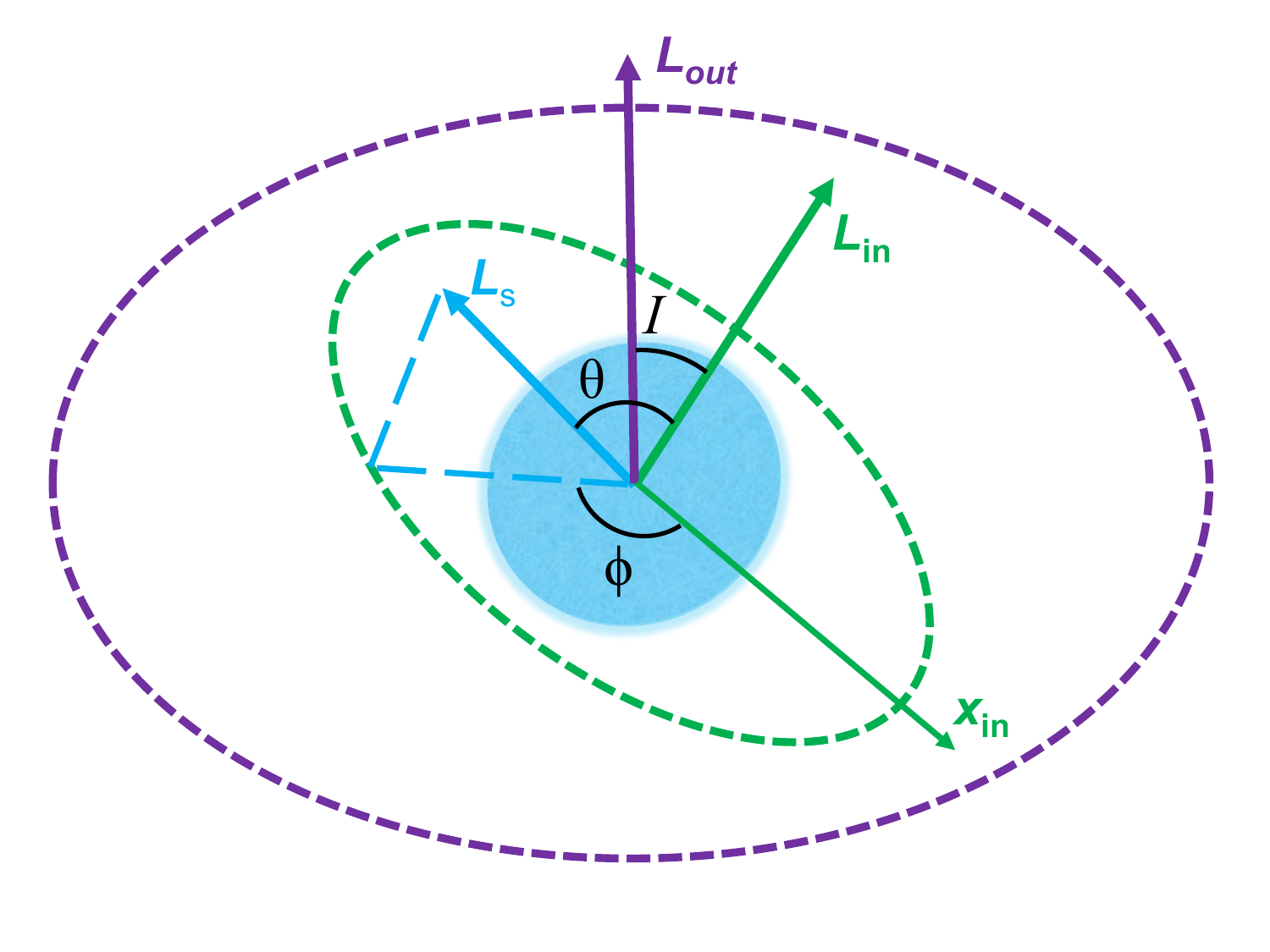}
\caption{Geometry in the adopted reference frame, with $z$-axis along the outer orbital angular momentum vector ${\bf L}_{\rm out}$, which is nearly constant. The reference frame co-rotates with the inner orbital angular momentum vector (${\bf L}_{\rm in}$) around ${\bf L}_{\rm out}$, with a constant orbital inclination $I$. The star's rotation axis is along ${\bf L}_s$, which precesses around ${\bf L}_{\rm in}$, with spin-orbit obliquity $\theta$ and precessional phase $\phi$.}
\label{fig:Cassini}
\end{figure}

We consider the situation where the spin angular momentum $L_s$ is much smaller than the inner orbital angular momentum $L_{\rm in}$, which in turn is much smaller than the outer orbital angular momentum $L_{\rm out}$, which is nearly constant. In this paper we consider only circular orbits. Figure \ref{fig:Cassini} illustrates the geometry of the system. The inner and outer orbital angular momenta are misaligned by the inclination angle $I$, while the spin and inner angular momentum vectors are misaligned by the angle $\theta$. The angle $\phi$ is the spin precession phase in the orbital plane.

Following \cite{Lai_2012} and \cite{Su_2021}, the spin rate $\Omega_s$ and misalignment angle $\theta$ evolve under the action of equilibrium tidal dissipation and orbital precession as
\begin{equation}
\label{eq:omsdot}
    \dot{\Omega}_s = \frac{\Omega_s}{t_s} \bigg[ \frac{2 n}{\Omega_s} \cos \theta - (1 + \cos^2 \theta) \bigg] \, ,
\end{equation}
\begin{equation}
\label{eq:thetadot}
    \dot{\theta} = -g \sin I \sin \phi - \frac{\sin \theta}{t_s} \bigg[ \frac{2 n}{\Omega_s} - \cos \theta \bigg] \, .
\end{equation}
Here, $n$ is the inner orbital frequency, and $g$ is the inner orbital precession frequency 
\begin{equation}
\label{eq:g}
    g = - \frac{3}{4} \cos I \, \frac{M_{\rm out}}{M_1+M_2} \bigg(\frac{a_{\rm in}}{a_{\rm out}}\bigg)^3 n \, ,
\end{equation}
where $M_1$ is the mass of the primary, $M_2$ is the secondary mass in the inner orbit, $M_{\rm out}$ is the tertiary msas on the outer orbit, and $a_{\rm in}$ and $a_{\rm out}$ are the corresponding semi-major axes. The orbital precession is similar to that which occurs during Kozai-Lidov cycles \citep{kozai:62,lidov:62,Naoz_2016}, though simpler since we consider only circular orbits.

The tidal synchronization rate is
\begin{equation}
 t_s^{-1} = \frac{3 k_2}{4 k Q} \frac{M_2}{M_1} \bigg(\frac{R_1}{a_{\rm in}}\bigg)^3 n \, .
\end{equation}
where $Q$ is the tidal quality factor, $k_2$ is the star's Love number, and its moment of inertia is $I = k M_1 R_1^2$, with $R_1$ the primary star's radius. The precessional angle $\phi$ changes due to both spin-orbit and orbital precession as
\begin{equation}
\label{eq:phidot}
    \dot{\phi} = - \alpha \cos \theta - g \big( \cos I + \sin I \cot \theta \cos \phi \big) \, ,
\end{equation}
where the spin-orbit precession frequency is 
\begin{equation}
\label{eq:alpha}
    \alpha = \frac{k_2}{2 k} \frac{M_2}{M_1} \bigg(\frac{R_1}{a_{\rm in}}\bigg)^3 \Omega_s \, .
\end{equation}

As discussed in prior works (e.g., \citealt{Su_2021}), there are one or two stable Cassini equilibria of these equations. For slow dissipation with $t_s^{-1} \ll g$, Cassini State 1 (CS1) has $\phi \simeq 0$, while Cassini State 2 (CS2) has $\phi \simeq \pi$. The spin equilibrium occurs when
\begin{equation}
\label{eq:omseq}
    \Omega_{s,{\rm eq}} \simeq \frac{2 n \cos \theta}{1 + \cos^2 \theta} \, .
\end{equation}
Obliquity equilibrium can be maintained when the spin-orbit precession is equal to orbital precession (equation \ref{eq:phidot}). In the limit that $\eta = -g/\alpha \ll 1$, CS1 has $\theta_{\rm eq} \simeq 0$ and $\Omega_{s,{\rm eq}} \simeq n$, i.e., nearly aligned and synchronous rotation. CS2 has $\cos \theta_{\rm eq} \simeq \sqrt{\eta_{\rm sync} \cos I/2}$ and $\Omega_{s,{\rm eq}} \simeq \sqrt{2 \eta_{\rm sync} \cos I}$, where
\begin{equation}
    \eta_{\rm sync} = \frac{3 k}{k_2} \frac{ M_1 M_{\rm out}}{M_2 (M_1+M_2)} \bigg(\frac{a_{\rm in}}{R_1}\bigg)^3 \bigg(\frac{a_{\rm in}}{a_{\rm out}}\bigg)^3 \cos I \, .
\end{equation}
Hence, when $\eta_{\rm sync} \ll 1$ as often occurs for both stellar and planetary orbital configurations, a star trapped in CS2 will have a nearly perpendicular obliquity and a very subsynchronous rotation rate.

\subsection{Evolution of Inner Orbit}

Because the spin and orbit remain misaligned and asynchronous in CS2, there is ongoing energy dissipation, which causes continued evolution of the inner orbit. This will typically cause orbital decay such that the orbital precession rate $g$ decreases, while the spin synchronization rate $t_s^{-1}$ increases. As discussed in \cite{Su_2021}, CS2 becomes unstable when the precession term in equation \ref{eq:thetadot} becomes smaller than the spin alignment term, which occurs when $t_s^{-1} > t_{s,c}^{-1}$, where
\begin{equation}
\label{eq:tsc}
    t_{s,c} = \frac{\sin \theta}{g \sin I} \bigg( \frac{2 n}{\Omega_s} - \cos \theta \bigg) \, .
\end{equation}
Assuming $\eta_{\rm sync} \ll 1$, this implies a minimum inner orbital period of
\begin{equation}
\label{eq:pmin}
    P_{\rm in,min} \approx 2 \pi \bigg( \frac{k_2 M_2 a_{\rm out}^3 R_1^3 }{G^2 k M_1 M_{\rm out}(M_1+M_2)} \bigg)^{1/4} \bigg( \frac{ 1}{Q \cos^2 I \sin I} \bigg)^{1/6} \, .
\end{equation}

The semi-major axis of the inner orbit evolves due to tidal dissipation as
\begin{equation}
\label{eq:adot}
    \frac{\dot{a}_{\rm in}}{a_{\rm in}} = - \frac{4}{t_s} \frac{I_s}{L_{\rm in}} \big( n - \Omega_s \cos \theta \big) \, .
\end{equation}
The semi-major axis evolves on a timescale longer than the characteristic spin evolution time, $t_{\rm in} = a_{\rm in}/\dot{a}_{\rm in} \sim (\Omega_s L_{\rm in}/ n L_s) t_s$ while the system is in CS2.

Dissipation also changes the mutual orbital inclination. The spin-orbit inclination angle is related to the orbital angular momenta via
\begin{equation}
\label{eq:idef}
   {\bf L} \cdot {\bf L} = L^2 = L_{\rm in}^2 + L_{\rm out}^2 + 2 L_{\rm in} L_{\rm out} \cos I \, .
\end{equation}
The inner orbit loses angular momentum to tidal torques on one (or both) stars but the outer orbit conserves its angular momentum because it only induces precession of the inner orbit (which does not change $L_{\rm out}$ or $I$). Taking a time derivative of equation \ref{eq:idef} then yields
\begin{equation}
\label{eq:cosIdot}
    \frac{d \cos I}{dt} = \frac{1}{L_{\rm in} L_{\rm out}} \bigg[ L \dot{L} - L_{\rm in} \dot{L}_{\rm in} - L_{\rm out} \dot{L}_{\rm in} \cos I \, \big] .
\end{equation}

Following \cite{Lai_2012}, the tidal torque on the orbit is opposite that of the star, hence for equilibrium tides
\begin{equation}
    \label{eq:Linvecdot}
    {\bf \dot{L}_{\rm in}} = \frac{I_s}{t_s} \bigg[ - \sin \theta \Omega_{\rm s} \hat{{\bf x}}_{\rm in} - 2 \big(n -  \Omega_{\rm s} \cos \theta \big) \hat{{\bf L}}_{\rm in} \bigg] \, . 
\end{equation}
Here, $\hat{{\bf x}}_{\rm in}$ and $\hat{{\bf L}}_{\rm in}$ specify the inner orbit's orientation (Figure \ref{fig:Cassini}). The inner orbit loses AM at the rate
\begin{equation}
\label{eq:Lindot}
    \dot{L}_{\rm in} = {\bf \dot{L}}_{\rm in} \cdot \hat{{\bf L}}_{\rm in} = - \frac{2 I_s}{t_s} \big(n - \Omega_{\rm s} \cos \theta \big) \, .
\end{equation}
The change in the total orbital angular momentum is
\begin{equation}
    \dot{L} \simeq {\bf \dot{L}}_{\rm in} \cdot \hat{{\bf L}}_{\rm out}
\end{equation}
Using $\hat{\bf{x}}_{\rm in} \cdot \hat{{\bf L}}_{\rm out} = \cos \phi \sin I$, $\hat{\bf{y}}_{\rm in} \cdot \hat{{\bf L}}_{\rm out} = \sin \phi \sin I$, and $\hat{\bf{z}}_{\rm in} \cdot \hat{{\bf L}}_{\rm out} = \cos I$, we have
\begin{equation}
\label{eq:Ldot}
    \dot{L} \simeq \frac{I_s}{t_s} \Big[ - \sin \theta \cos \phi \sin I \Omega_s - 2  \cos I (n - \Omega_s \cos \theta) \Big] \, .
\end{equation}

There is also a non-dissipative spin-orbit precession term that enters at the same order of $L_s/L_{\rm in}$. Conservation of AM requires spin-orbit precession to cause ${\bf \dot{L}}_{\rm in} = - {\bf \dot{L}}_{\rm s} \simeq \alpha L_s \sin \theta \cos \theta \hat{\bf{y}}_{\rm in}$, and hence 
\begin{equation}
\label{eq:Ldotspinorbit}
   \dot{L} = \alpha L_s \sin \theta \cos \theta \sin \phi \sin I. 
\end{equation}
Inserting equation \ref{eq:Lindot} and both contributions to $\dot{L}$ (equations \ref{eq:Ldot} and \ref{eq:Ldotspinorbit}) into equation \ref{eq:cosIdot}, we find
\begin{equation}
\label{eq:idot}
    \dot{I} \simeq  \frac{L_s}{L_{\rm in}} \bigg( \frac{\sin \theta \cos \phi}{t_s} - \sin \theta \cos \theta \sin \phi \, \alpha \bigg) \, .
\end{equation}
Here, we have ignored the $L \dot{L}_{\rm in}$ term because it is smaller by a factor of $L_{\rm in}/L_{\rm out}$. We note that the non-dissipative term in equation \ref{eq:idot} is identical to that found by \cite{Anderson_2018}, using their equation 20 and plugging in the non-dissaptive term in equation \ref{eq:thetadot}. While this term appears to be larger than the dissipative term if $\alpha > t_s^{-1}$, it becomes small in a Cassini state where $\sin \phi \simeq 0$, and so both terms should be included.



In CS1 where $\phi \simeq 0$, tidal dissipation causes the orbit evolve to evolve towards larger values of $I$, i.e., away from alignment, while the opposite is true of CS2 where $\phi \simeq \pi$. This can be seen in Figure \ref{fig:Cassini}, remembering that the tidal torque drives the inner orbit's AM towards that of the star's spin vector.

If initially $I > \pi/2$, a system in CS2 will evolve towards $I = \pi/2$, i.e., towards orbits that are perpendicular to each other. As $I \rightarrow \pi/2$, so too $\theta \rightarrow \pi/2$ in CS2. Hence, the system evolves to a state where the spin AM of the star becomes aligned with the \textit{outer} orbit, while the inner orbit is perpendicular to both the star's spin and the outer orbit. At this stage, the rate of evolution approaches zero because $\Omega_s \sim 2 \cos \theta \rightarrow 0$. Hence, although $I \simeq \theta \simeq \pi/2$ is not a stable equilibrium, a system could spend a very long time evolving through this state and could be likely to be observed in this state. In this case, the star's spin rate would be nearly zero, slower than the CS2 equilibrium at a random mutual orbital inclination.

However, this scenario is unlikely to occur in real systems because the inclination angle evolves slowly compared to the semi-major axis while in CS2 where $\Omega_s \ll n$. Hence, the system must be fine-tuned, with initial $I$ just slightly larger than $\pi/2$, in order for this to happen. Instead, we shall see that orbital decay usually causes the system to be disrupted from CS2 before it reaches $I \simeq \pi/2$.

\section{Cassini States with Inertial Wave Dissipation}
\label{sec:inertial}

Above and in most prior work, tidal dissipation was modeled via the equilibrium tide assumption. Real tidal dissipation often occurs via dynamical tides, which can be very sensitive to the tidal forcing frequency. Hence, some components of tidal forcing can cause more dissipation than others, modifying the spin/orbital evolution. As discussed in \cite{Lai_2012}, the component of tidal dissipation due to $m=\pm1$ waves in the star forced by the $m'=0$ component of the tidal potential have a frequency $\omega_{m m'} = m' n - m \Omega_s = \mp \Omega_s$ in the star's rotating frame. This allows for resonance with inertial waves, which could cause more tidal dissipation than equilibrium tides or gravity waves.

\begin{figure*}
\centering
\includegraphics[width=0.49\textwidth]{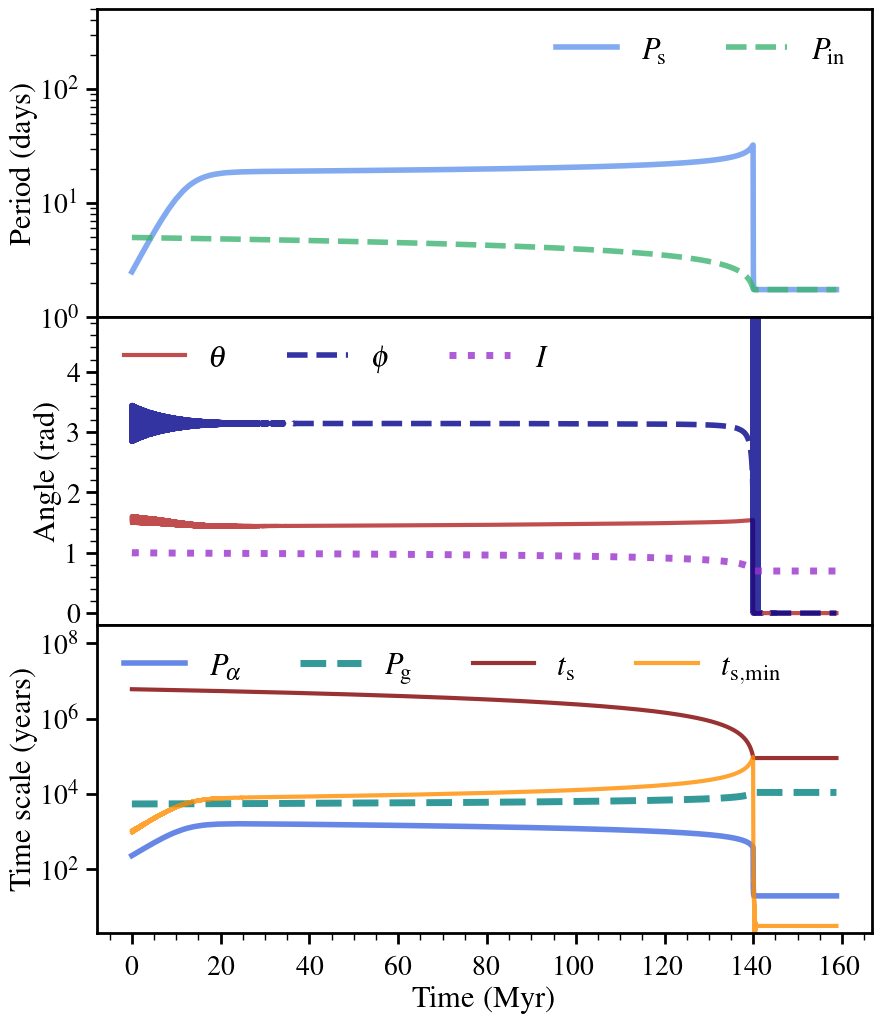}
\includegraphics[width=0.49\textwidth]{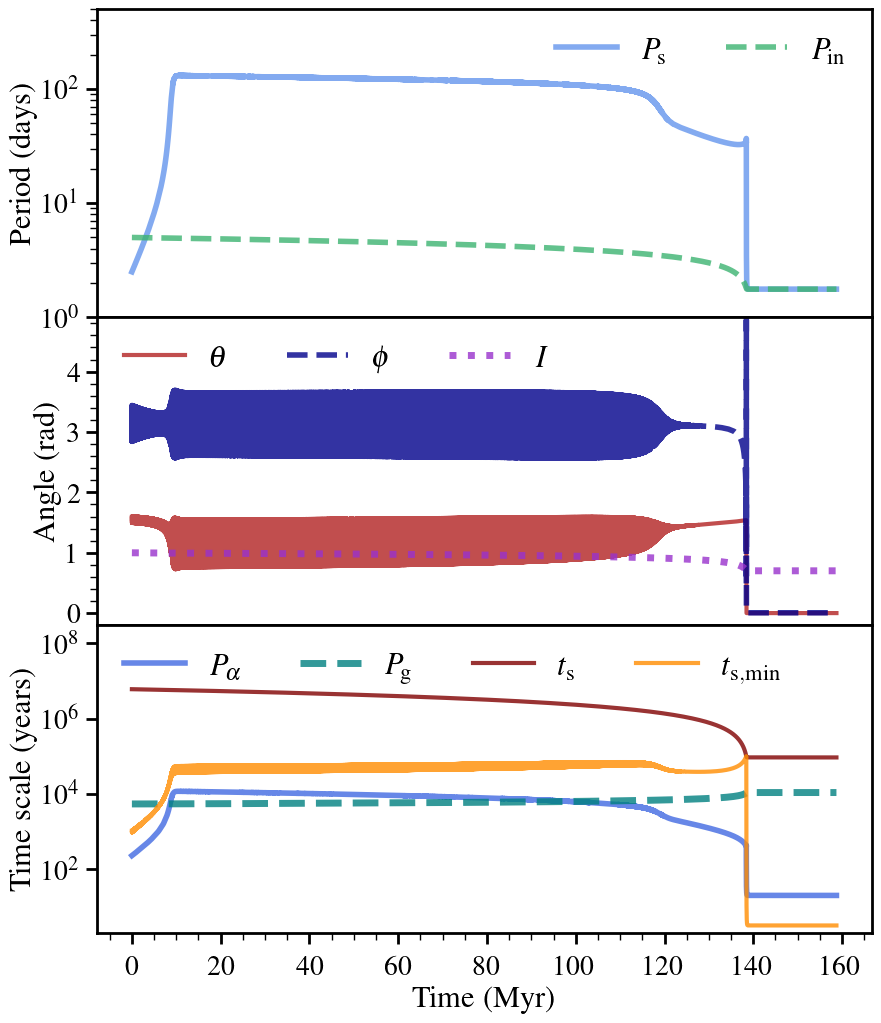}
\caption{ {\bf Left:} Spin and orbital evolution as a function of time for an equilibrium tidal model. {\bf Right:} Evolution for a model with extra inertial wave dissipation with $X_{10} = 300$. {\bf Top row:} The star's spin period $P_{\rm s}$ and inner orbital period $P_{\rm in}$. {\bf 2nd row:} Mutual orbital inclination angle $I$, spin-orbit obliquity $\theta$, and precessional phase $\phi$.
{\bf Bottom row:} Spin-orbit precession period $P_{\alpha}$, orbital precession period $P_{\rm g}$, tidal synchronization time scale $t_{\rm s}$, and minimum synchronization time scale $t_{\rm s,min}$. In both cases, the star reaches its CS2 equilibrium after $\sim$10 Myr, which destabilizes at $\sim$140 Myr and the star moves to a nearly synchronous and aligned CS1 state.}
\label{fig:SlowSpinEvol}
\end{figure*}

We specify the importance of ``extra" inertial wave dissipation via a parameter $X_{10} = \tau_{10}/\tau - 1$, where $\tau_{10}/\tau$ is the dissipation efficiency of inertial waves relative to equilibrium tides. The value of $X_{10} \rightarrow 0$ in the absence of inertial wave dissipation, but it can become very large for efficient inertial wave dissipation. The orbital evolution equations are modified to 
\begin{equation}
\label{eq:omsdot10}
    \dot{\Omega}_s = \frac{\Omega_s}{t_s} \bigg[ \frac{2 n}{\Omega_s} - (1 + \cos^2 \theta) -\frac{X_{10}}{2} \sin^2 \theta \cos^2 \theta \bigg] \, ,
\end{equation}
\begin{equation}
\label{eq:thetadot10}
    \dot{\theta} = -g \sin I \sin \phi - \frac{\sin \theta}{t_s} \bigg[ \frac{2 n}{\Omega_s} - \cos \theta + \frac{X_{10}}{2} \cos^3 \theta \bigg] \, ,
\end{equation}
\begin{equation}
\label{eq:idot10}
    \dot{I} \simeq  \frac{L_s}{L_{\rm in}} \bigg( \frac{\sin \theta \cos \phi}{t_s} \bigg[ 1 + \frac{X_{10}}{2} \cos^2 \theta \bigg] - \sin \theta \cos \theta \sin \phi \, \alpha \bigg) \, .
\end{equation}
The equations for $\dot{\phi}$ and $\dot{a}_{\rm in}$ are unmodified, since inertial wave dissipation does not cause orbital decay as discussed in \cite{Lai_2012}.

The Cassini equilibria will be altered by inertial wave dissipation. The new spin equilibrium is
\begin{equation}
    \Omega_{s,{\rm eq}} = \frac{4 \cos \theta \, n}{2 + 2 \cos^2 \theta + X_{10} \sin^2 \theta \cos^2 \theta } \, .
\end{equation}
In the limit of large inertial wave dissipation with $X_{10} \gg 1$, we have
\begin{equation}
\label{eq:oms10}
    \Omega_{s,{\rm eq}} \simeq \frac{4 \, n}{X_{10} \sin^2 \theta \cos \theta } \, .
\end{equation}
Cassini equilibrium occurs where $\dot{\phi}\simeq0$, requiring
\begin{equation}
    \alpha \cos \theta = g \big( \cos I + \sin I \cot \theta \cos \phi \big) \, .
\end{equation}
Plugging in equation \ref{eq:oms10} and using equation \ref{eq:alpha}, we have
\begin{equation}
\label{eq:inertialeq}
X_{10} \eta_{\rm sync} \big( \cos I \sin \theta + \sin I \cos \theta  \cos \phi \big) \sin \theta \simeq 4  \, .
\end{equation}
If $X_{10} \eta_{\rm sync} \ll 1$, then there is no solution to equation \ref{eq:inertialeq}, and instead CS2 will be somewhat modified from its equilibrium tide value, as discussed in Felce \& Fuller in prep.

However, if $X_{10} \eta_{\rm sync} \gg 1$ and the system is in a Cassini state with $\cos \phi \simeq \pm 1$, then the solution must have $\theta \simeq \pi - I$ or $\theta \simeq I$. The corresponding spin rate is
\begin{equation}
\label{eq:omseq10}
    \Omega_{s,{\rm eq}} \simeq \mp \frac{4 n}{X_{10} \sin^2 I \cos I} \, .
\end{equation}
Here, the $-$ sign corresponds to CS1 ($\phi\simeq0$) whose equilibrium only exists for $\pi/2 < I < \pi$, while the $+$ sign corresponds to CS2 ($\phi \simeq \pi$) whose equilibrium only exists for $0 < I < \pi/2$.
Interestingly, the rotation rate can be very slow in this state, much slower than in the absence of inertial wave dissipation. Furthermore, the new equilibrium is independent of the radius of the star, and the outer companion's mass or semi-major axis. It is dependent only on the value of $X_{10}$ and the mutual orbital inclination. In CS2, the spin vector of the star is nearly aligned with the outer orbit, while in CS1 it is nearly anti-aligned.

\section{Orbital Evolution Calculations}
\label{sec:orb}

\begin{table*}
\caption{Properties of five stellar systems with very slowly rotating stars in close binaries known to be part of triple systems. The columns indicate the name, primary mass, companion mass, primary radius, inner orbital period, primary rotation period, and outer orbital period. Reference labels are 1: \citealt{Lampens_2018}, 2: \citealt{Gaia_DR2}, 3: \citealt{Li_2020a}, 4: \citealt{Zhang_2019}, 5: \citealt{Borkovits_2016}, 6: \citealt{Windemuth_2019}, 7: \citealt{Zhang_2020}, 8:
\citealt{sekaran:20}, 9:
\citealt{sekaran:21}, 10: 
\citealt{Fekel_2019}, 11: 
\citealt{Kallinger_2017}.}
\hskip-2.0cm \begin{tabular}{@{}cccccccc@{}}
\toprule
ID & $\textbf{M}_{\textbf{1}} \, (M_\odot)$ & $\textbf{M}_{\textbf{2}} \, (M_\odot)$ & $\textbf{R}_{\textbf{1}} \, (R_\odot)$ & $\textbf{P}_{\rm orb} \, ({\rm d})$ & $\textbf{P}_{\rm rot} \, ({\rm d})$ & $\textbf{P}_{\rm out} \, ({\rm d})$ & $\textbf{Refs}$\\
\midrule
KIC 4480321 & $1.5^{+0.3}_{-0.2} $& $1.5^{+0.3}_{-0.2} $& $1.9^{+0.5}_{-0.5} $& $9.166^{+6e-05}_{-6e-05} $& $121.0^{+4.0}_{-4.0} $& $2280.0^{+29.0}_{-29.0} $& 1, 2, 3\\
KIC 8429450 & $1.68^{+0.2}_{-0.13} $& $1.462^{+0.174}_{-0.113} $& $2.438^{+0.083}_{-0.081} $& $2.705^{+2e-07}_{-2e-07} $& $38.0^{+128.0}_{-17.0} $& $3088.0^{+1700.0}_{-1700.0} $& 3, 4, 5, 6\\
KIC 9850387& $1.66^{+0.01}_{-0.01} $& $1.062^{+0.003}_{-0.005} $& $2.154^{+0.002}_{-0.004} $& $2.748^{+7e-07}_{-4e-07} $& $188.68^{+74.5}_{-41.6} $& $671.0^{+2.0}_{-2.0} $& 3, 5, 7, 8, 9\\
HD 126516 & $1.34^{+0.2}_{-0.2} $& $0.28^{+0.03}_{-0.03} $& $1.66^{+0.08}_{-0.08} $& $2.124^{+1e-07}_{-1e-07} $& $18.3^{+2.8}_{-7.7} $& $702.71^{+0.25}_{-0.25} $& 10\\
HD 201433 & $3.05^{+0.025}_{-0.025} $& $0.7^{+0.3}_{-0.3} $& $2.6^{+0.2}_{-0.2} $& $3.313^{+5e-04}_{-5e-04} $& $292.0^{+76.0}_{-76.0} $& $154.2^{+0.03}_{-0.03} $& 11\\
\bottomrule
\end{tabular}
\label{tab:system_params}
\end{table*}

To examine the spin and orbital evolution of a realistic stellar system, we numerically integrate equations \ref{eq:omsdot}, \ref{eq:thetadot}, \ref{eq:phidot}, \ref{eq:adot}, and \ref{eq:idot} forward in time. For simplicity, we ignore evolution of the star's structure and radius. We choose initial conditions $P_{\rm in} = 5$ days, $P_{\rm out} = 10^3$ days, $I = 1$, and use a tidal lag time $\tau=10^{-3} \, {\rm s}$, which corresponds to an initial tidal quality factor of $Q \sim 10^6$. The system has $M_1 = 2 \, M_\odot$, $M_2 = M_\odot$, $M_{\rm out} = M_\odot$, $R_1= 2 R_\odot$, $k_2 = 0.003$, and $k=0.036$. We set the initial spin rate $\Omega_s = 2 n$, obliquity $\theta = \pi/2$, and precession angle $\phi = \pi$. This allows the system to become trapped in CS2, though the probability of this occurring depends on the initial conditions as discussed in \cite{Su_2021}.

\begin{figure*}
\centering
\includegraphics[width=0.49\textwidth]{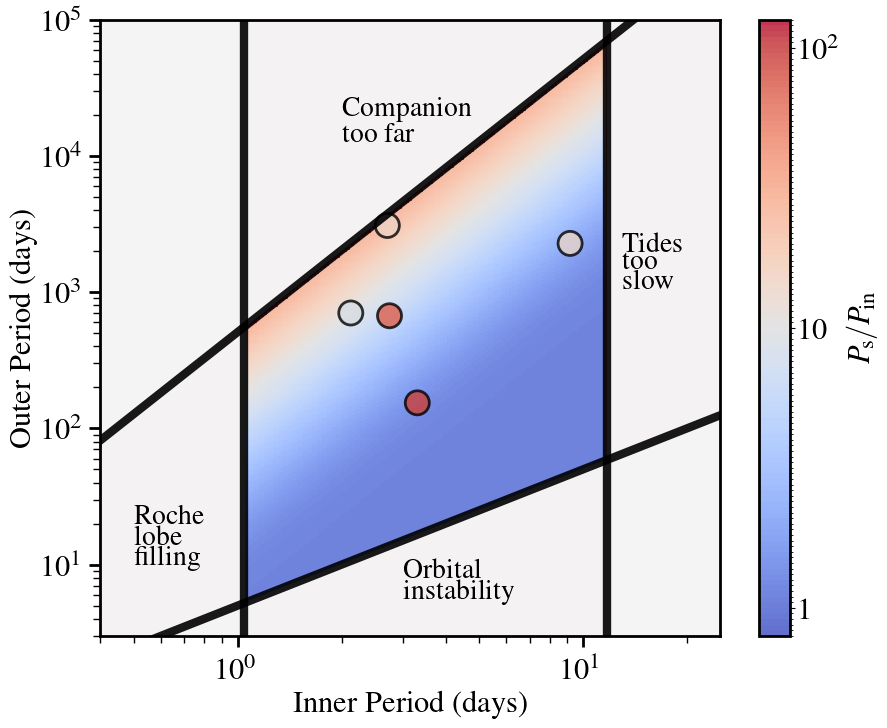}
\includegraphics[width=0.49\textwidth]{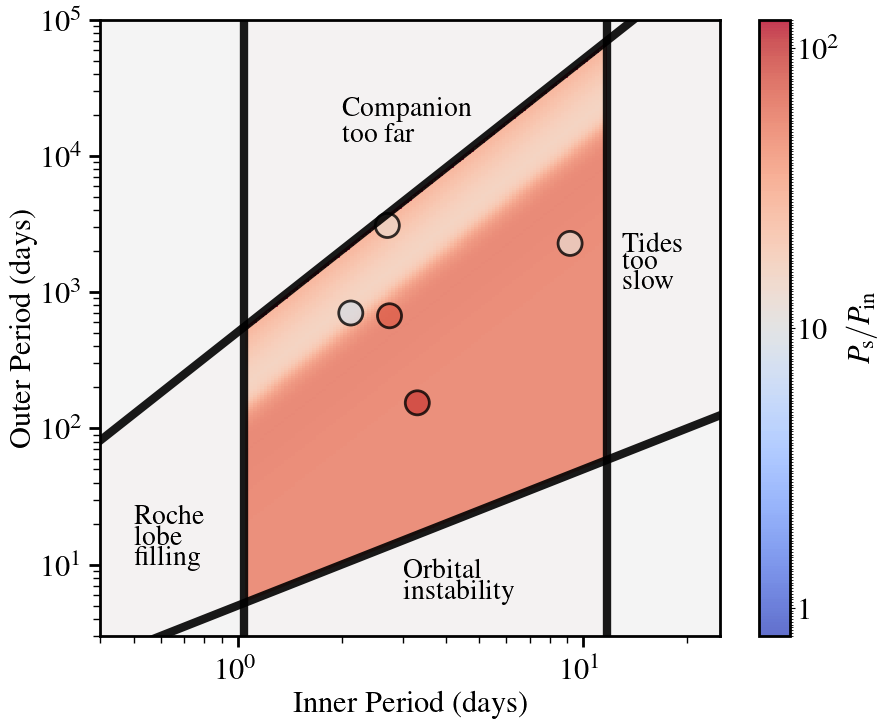}
\caption{
Period range in which systems can be trapped in Cassini State 2, for a $2 \, M_\odot$ star and $1 \, M_\odot$ companions. Color shading indicates the equilibrium rotation period $P_s$ relative to the inner orbital period $P_{\rm in}$.
Colored points are observed systems in Table \ref{tab:system_params}. {\bf Left:} Predicted spin period assuming equilibrium tidal dissipation. {\bf Right:} Predicted spin period including extra inertial wave dissipation with $X_{10}=600$.}
\label{fig:Prange}
\end{figure*}

The left panel of Figure \ref{fig:SlowSpinEvol} shows the resulting evolution. The system quickly becomes trapped in CS2 such that $\phi$ and $\theta$ librate around their equilibrium positions of $\phi_{\rm eq} \simeq \pi$ and $\theta_{\rm eq} \simeq \pi/2$. On a time scale of $t_s \! \sim \! 10^7 \, {\rm yr}$, the spin period approaches its equilibrium value as predicted by equation \ref{eq:omseq}, which is a factor $\sim$10 times longer than the inner orbital period. On a time scale of $(\Omega_s L_{\rm in}/n S) t_s \sim 10^8 \, {\rm yr}$ the orbital period decreases in response to ongoing tidal dissipation. At $t\simeq 140$Myr, CS2 becomes unstable because $t_s < t_{s,c}$, and the system very rapidly evolves into CS1, where the star's spin becomes nearly synchronized and aligned with the inner orbit. Tidal dissipation essentially ceases, and the orbital period becomes nearly constant. 

We now examine the evolution with added inertial wave dissipation. Equations $\dot{\Omega}_s$, $\dot{\theta}$, and $\dot{I}$ are replaced by equations \ref{eq:omsdot10}, \ref{eq:thetadot10}, and \ref{eq:idot10}, using an enhanced inertial wave dissipation rate of $X_{10} = 300$. The right panels of Figure \ref{fig:SlowSpinEvol} show the resulting evolution. As before, the system quickly becomes trapped in CS2 such that $\phi$ and $\theta$ librate around their equilibrium positions. Now, however, $\theta_{\rm eq} \sim I \simeq 1$ rather than $\approx \pi/2$ as it did without inertial waves. The libration around the equilibrium values of $\theta_{\rm eq}$ and $\phi_{\rm eq}$ is also much larger in this case. Most importantly, the equilibrium spin rate is much smaller, as predicted by equation \ref{eq:omseq10}.

After $\approx$115 Myr, $\eta_{\rm sync} X_{10} \lesssim 1$ and the analytic solution of Section \ref{sec:inertial} breaks down. The system transitions to a different equilibrium similar to the ordinary CS2 obtained without inertial wave dissipation, resulting in a faster spin rate. Throughout the evolution, the inner orbital period decays at roughly the same rate as it did without inertial waves, because they do not cause orbital decay. The critical tidal dissipation time scale $t_{s,c}$ of equation \ref{eq:tsc} is only moderately changed by the extra tidal dissipation from inertial waves. Hence, CS2 is destabilized after $\approx$135 Myr years when $t_s \lesssim t_{s,c}$, at which point the system approaches a nearly synchronous and aligned state in CS1.

\vspace{-10pt}
\section{Discussion}

\subsection{Comparison with Observed Systems}

We have shown that stars in triple systems can get caught in CS2, in which one (or both) stars of the inner binary has a misaligned and potentially very sub-synchronous rotation rate. To get a sense of the parameter range over which we expect stellar triples to possibly be caught in CS2, we compute the expected spin period of the inner primary as a function of inner and outer orbital period. We use the same parameters as Section \ref{sec:orb}, but with $I = \pi/4$, and a tidal quality factor $Q= 2 \times 10^5$.
We compute the CS2 equilibrium by numerically solving for the equilibrium of equations \ref{eq:omsdot}, \ref{eq:thetadot}, and \ref{eq:phidot} (see method discussed in \citealt{Fabrycky_2007}).

Figure \ref{fig:Prange} shows the expected spin periods for stars that could be trapped in CS2. Inner orbital periods less than $\sim$1 day imply Roche lobe overflow, while for orbital periods greater than $\sim$10 days, the synchronization time scale is longer than the main sequence lifetime. Orbits with $P_{\rm out} \lesssim 5 P_{\rm in}$ are unstable \citep{Holman_1999}. Finally, CS2 is destabilized for outer orbits that are too distant, as defined by equation \ref{eq:pmin}. Stronger tides or larger stellar radii would increase the limiting orbital periods (i.e., the left/right/upper boundaries of the plot would shift to the right). 

Figure \ref{fig:Prange} shows that the spin period of the inner star is expected to lie in the range $P_{\rm s} \sim 1-30 P_{\rm in}$ for this choice of stellar parameters. Closer third bodies produce faster orbital precession, larger $\eta_{\rm sync}$, and faster rotation in CS2. Distant third bodies near the boundary of CS2 stability can lead to much slower stellar rotation.

The five well-characterized triple systems that harbor very slowly rotating primaries (Table \ref{tab:system_params}) are also shown in Figure \ref{fig:Prange}. Two of these (KIC 8429450 and HD 126516) have spin periods similar to those predicted by equilibrium tidal theory. However, the other three systems rotate much slower than predicted. While the masses and radii of each system is different from the values used to make the figure, using appropriate values still leads to predicted rotation rates that are too large.

The right panel of Figure \ref{fig:Prange} shows the predicted rotation rates accounting for enhanced inertial wave dissipation with $X_{10}=600$. Much smaller spin rates (equation \ref{eq:omseq10}) are expected when $X_{10} \eta_{\rm sync} \gtrsim 1$, as produced in the lower-right red part of the figure. These predictions come closer to matching the observed systems, especially the two reddest points (HD 201433 and KIC 9850387). Hence the very slow rotation of those stars may be evidence for tidal dissipation via inertial waves. If so, the observed spin period tells us the effective value of the inertial wave dissipation, with $X_{10} \sim 10^3$ needed to match the data. Future work should calculate $X_{10}$ from first principles to determine whether inertial wave dissipation can explain the observed slow rotation rates. 

Figure \ref{fig:Prange} also shows that relatively compact triples (outer orbital periods less than $\sim$100 yr, or semi-major axes less than $\sim$30 AU for Sun-like stars) are likely required for stars to be in CS2. At these separations, observations indicate that the mutual inclinations of the two orbits are typically quite small, with average inclination angles of $I \sim 20^\circ$ \citep{tokovinin:17}. Small inclination angles increase the minimum inner orbital period (equation \ref{eq:pmin}) and increase the equilibrium spin rate for inertial wave dissipation (equation \ref{eq:omseq10}). The small misalignment angles of compact triples may thus be important for understanding the population of systems in CS2. 

\subsection{Predicting Tertiary Companions}

If slowly rotating stars in close binaries are trapped in CS2, they can be used to infer the presence of a third body and predict its orbital separation. For instance, using a measured spin period and inner binary orbital period allows one to predict the outer orbital period given a tidal model, as discussed in Felce \& Fuller, in prep. Predictions for the tertiary are uncertain because the CS2 equilibrium depends on the tidal dissipation physics, but typically we expect outer orbital periods of $\sim \! 10^2-10^5 \, {\rm d}$ for main sequence primary stars.

Misaligned stars in close binaries could also be caught in CS2 and signal the presence of a tertiary companion. Two well known binary systems DI Her \citep{Albrecht_2009} and CV Vel \citep{Albrecht_2014} have primary spin axes misaligned with the orbit, possibly indicating they are in CS2. However, both stars in DI Her rotate faster than the orbit, disfavoring the CS2 hypothesis. In CV Vel, the rotation rate is comparable to the orbital frequency. If caught in CS2, this suggests a nearby tertiary with orbital periods of $\sim$hundreds of days (i.e., the dark blue region of the left panel of Figure \ref{fig:Prange}), which could be tested with radial velocity monitoring. However, it is also possible that this system is a binary that is still approaching tidal synchronization and alignment, as proposed by \cite{Albrecht_2014}.

\subsection{Missing Physics}

We have discussed the impact of inertial wave dissipation on spin rates and obliquities of stars in CS2, but there are other effects that could be important. Low-mass stars undergo spin-down via magnetic braking, producing an additional term in equation \ref{eq:omsdot} that alters the CS2 equilibrium, as discussed in Felce \& Fuller 2023. However, magnetic braking is unlikely to greatly affect the intermediate-mass star systems studied here. Other types of dynamical tidal dissipation, such as traveling gravity waves or standing gravity modes, will alter the spin-up and alignment torques and hence the locations of the Cassini states.

Another possibility is differential rotation: equation \ref{eq:alpha} and all of our calculations have assumed the primary star is rigidly rotating, but differential rotation could allow for altered dynamics. In both HD 201433 \citep{Kallinger_2017} and KIC 9850387 \citep{sekaran:21}, asteroseismic analyses indicate the surface rotates much faster than the core. The faster surface rotation could increase the spin-orbit precession rate or alter the tidal dissipation mechanism, impacting the Cassini states.

A third effect which could prove to be important is altered spin-orbit precession for very slowly rotating stars. If the stellar quadrupole moment is not dominated by centrifugal distortion but some other distortion (e.g., tidally excited oscillations), then the spin-orbit precession rate can be altered from that of equation \ref{eq:alpha}, again producing different Cassini equilibria. We hope to examine some of these effects in future work.

\section{Conclusion}

We have shown that stars in triple systems can become caught in a Cassini State (CS2) characterized by a high spin-orbit obliquity and slow rotation of one or both stars in the inner binary. The period range over which this can occur for main sequence stars is an inner binary period of $P_{\rm in} \! \sim \! 1-10 \, {\rm d}$ with tertiary periods of $P_{\rm out} \! \sim \! 10-10^5 \, {\rm d}$ (Figure \ref{fig:Prange}), depending on the masses, primary stellar radius, and mutual orbital inclination. Such systems would stand out as having very long rotation periods $P_s \sim 10-10^3 \, {\rm d}$ in contrast to the expectation of tidal synchronization at short orbital periods. Indeed, several such systems have already been discovered (Table \ref{tab:system_params}), all of which have inner and outer orbital periods in the right range. Hence, we believe it is very likely that these systems are caught in CS2 and can be used to study the relevant dynamics.

Some of the observed systems rotate much slower than predicted, indicating our understanding of the relevant dynamics or tidal physics is incomplete. We have shown that tidal dissipation via inertial waves can potentially explain the very slow rotation rates in those systems. If so, the rotation rates translate to direct constraints on the tidal torques produced by inertial waves. Stars in Cassini states thus offer an exciting new pathway to study the orbital dynamics and tidal dissipation of stellar systems, but more theoretical work is needed for reliable interpretation of measurements. We encourage additional monitoring of stars in binaries to measure spin periods and search for third bodies, allowing for new discoveries of stars in CS2. The five systems listed here are probably just the tip of the iceberg, with many more systems likely to be discovered upon closer investigation of existing and future data.

\section*{Acknowledgments}

We thank Yubo Su for useful discussions and feedback on this work. 

\section*{Data Availability}

The orbital integration code, and plotting scripts to make Figures 2 and 3, are available upon request.

\bibliography{bib}

\end{document}